%% file: template.tex
\documentclass[journal]{vgtc}                     

\onlineid{0}
\vgtccategory{Research}

\title{CritLens: Visual Analytics for Criteria Discovery in Review-Based Decision Making}

\author{%
  Hongjia Wu,
  Shuai Zhou,
  Hongxin Zhang, and
  Wei Chen
}

\authorfooter{
  \item Hongjia Wu is with State Key Lab of CAD\&CG, Zhejiang University. E-mail: hongjiawu@zju.edu.cn.
  \item Shuai Zhou is with State Key Lab of CAD\&CG, Zhejiang University. E-mail: zhoushua1@zju.edu.cn.
  \item HongXin Zhang is with State Key Lab of CAD\&CG, Zhejiang University. E-mail: zhx@zju.edu.cn.
  \item Wei Chen is with State Key Lab of CAD\&CG, Zhejiang University. E-mail: chenvis@zju.edu.cn.
}

\abstract{%
We present \sys{}, a visual analytics system that helps users build personalized multi-criteria decision models from review text. In everyday decisions --- choosing equipment, hotels, or restaurants --- evaluation criteria are either preset by platforms or generated by LLMs, leaving users unable to discover, adjust, or verify them against the underlying evidence. This is problematic because many preferences are latent: they surface only upon encountering specific reviews, and any fixed framework risks overlooking low-frequency but decisive details. \sys{} addresses this gap by using LLMs to transform reviews into an initial AHP decision model, then supporting iterative, human-in-the-loop refinement. Through coverage gap detection in the embedding space, users discover criteria missed by the initial model; through interactive weight adjustment under AHP consistency constraints, they express personal priorities; and through a multi-level scorecard and exportable decision report, they trace every ranking back to the original review text. Two case studies, an eight-participant user study, and a quantitative consistency-repair experiment demonstrate the system's effectiveness.
}

\keywords{Visual analytics, multi-criteria decision making, analytic hierarchy process, large language models, review text analysis}

\teaser{
  \centering
  \includegraphics[width=\linewidth]{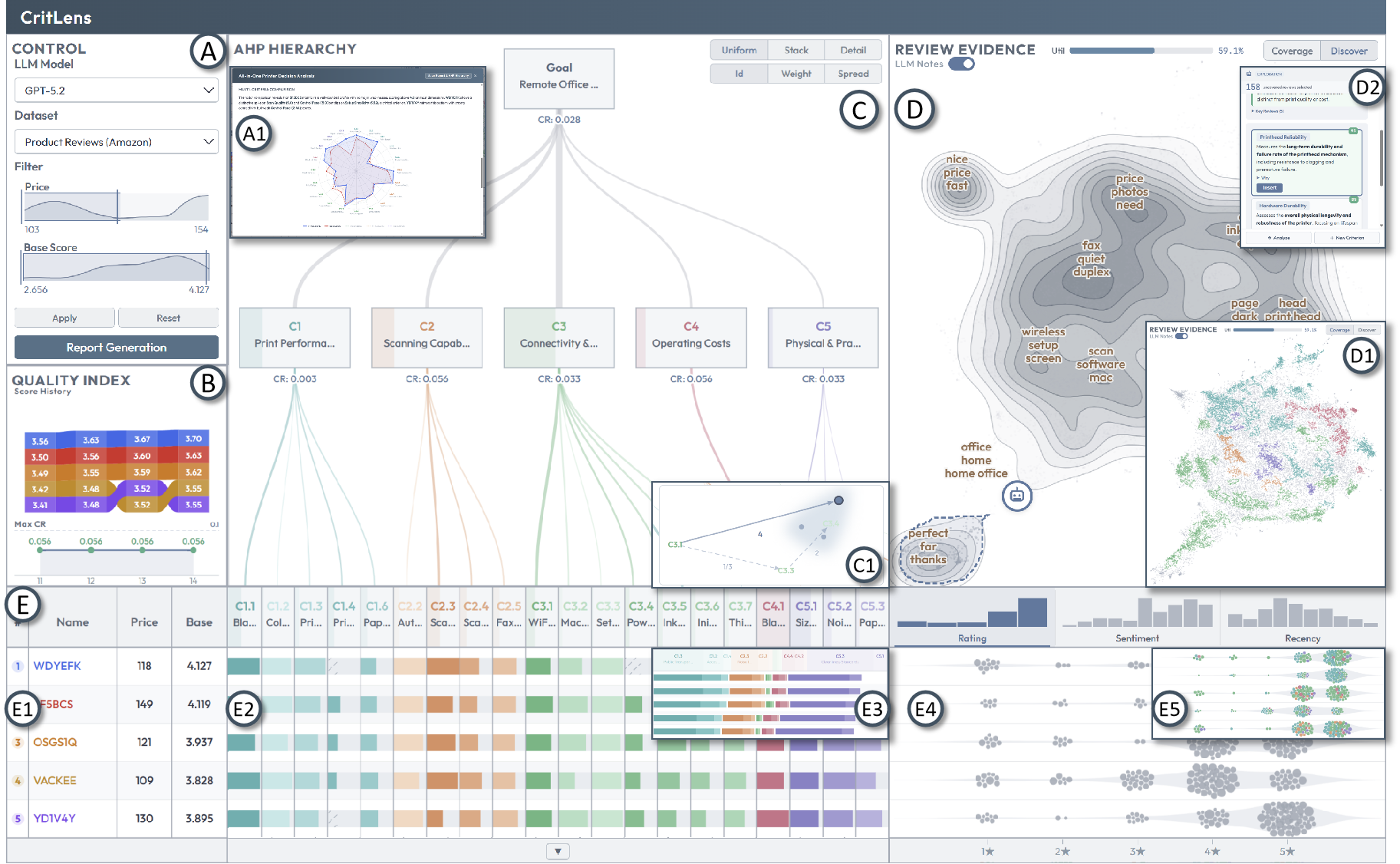}
  \caption{%
  The \sys{} visual analytics interface. (A)~Control Panel for dataset selection and price/rating filtering. (A$_1$)~Decision Report with ranking table, weight treemap, radar comparison, and evidence summary. (B)~Quality Index Panel: ribbon chart tracking score evolution of top-5 alternatives and line chart monitoring maximum consistency ratio. (C)~AHP Hierarchy View showing criteria hierarchy, weight allocation, and consistency status; (C$_1$)~Consistency Diagnosis View mapping pairwise comparison consistency via joint MDS projection. (D)~Review Evidence View projecting review snippets into a 2D semantic space with coverage and explore modes; (D$_1$)~coverage mode colors snippets by assigned criterion; (D$_2$)~LLM Analysis Panel displaying summary analysis and candidate criteria suggestions. (E)~Scorecard View presenting the full evidence chain per alternative: (E$_1$)~ranking and basic information; (E$_2$)~criterion score matrix with stacked color blocks; (E$_3$)~stacked mode scaling column widths by weight; (E$_4$)~beeswarm plot with kernel density overlay showing snippet distributions across rating/sentiment/recency; (E$_5$)~beeswarm in explore mode.%
  }
  \label{fig:teaser}
}

\graphicspath{{figs/}{figures/}{../paper/figures/}{./}}

\usepackage{amsmath,amssymb,amsthm}
\usepackage{booktabs}
\usepackage{enumitem}
\usepackage{mathptmx}

\newcommand{\sys}{CritLens}

\begin{document}


\firstsection{Introduction}

\maketitle
\label{sec:intro}

\input{body}

\acknowledgments{%
The authors wish to thank the domain experts for their valuable feedback throughout the design process.%
}

\bibliographystyle{abbrv-doi}
\bibliography{template}

\end{document}

%% file: body.tex

Purchasing office equipment for a team, choosing a restaurant for a corporate event, selecting a hotel before a business trip---these decisions all require multi-dimensional evaluation across many alternatives. Review text is a valuable source of authentic user feedback, yet evaluation criteria are either preset by platforms or generated as black-box outputs by large language models (LLMs), leaving decision-makers unable to participate in criteria discovery or inspect the evidence behind scores. Building a criteria framework tailored to specific needs from massive text poses three challenges.

\textbf{Latent preferences cannot be enumerated upfront.} Cognition during complex choices is nonlinear~\cite{pirolli2005sensemaking}; many preferences surface only after encountering specific text---for example, a procurement officer reading ``the driver broke completely after a macOS update'' suddenly realizes that OS compatibility is a key requirement. Criteria therefore cannot be defined all at once but must be constructed progressively during data exploration. \textbf{Static frameworks miss long-tail information.} Preset or auto-generated dimensions tend to capture high-frequency topics, while low-frequency details that may be decisive for a particular decision are easily overlooked. \textbf{Black-box aggregation lacks an evidence trail.} When a decision must be justified to others, decision-makers need to explain the basis of each dimension's score---how many reviews support it and what their quality and sentiment are---and conclusions without traceability cannot withstand scrutiny.

Interactive visual analytics offers a viable path~\cite{endert2012semantic}: machines handle massive text and compute structured models, while humans direct criteria selection, weight assignment, and evidence inspection---a division of labor consistent with the general framework of visual-analytics-assisted machine learning~\cite{sacha2017human,sacha2019vis4ml}. However, existing multi-attribute ranking visualization tools (e.g., LineUp~\cite{gratzl2013lineup}, WeightLifter~\cite{pajer2017weightlifter}) operate on predefined structured dimensions and do not address the discovery and construction of new dimensions from unstructured text; meanwhile, automated LLM methods lack interactive mechanisms for users to intervene in criteria restructuring.

We present \sys{}, an LLM-powered visual analytics system for text-driven multi-criteria decision making. The system uses LLMs to transform unstructured reviews into a structured Analytic Hierarchy Process (AHP) decision model and supports users in discovering new dimensions, inspecting scoring evidence, iteratively adjusting weights, and generating decision reports that link rankings back to the original review evidence. \sys{} targets domain experts and professional consumers who are accountable for their decisions. Our main contributions are:

\begin{itemize}[leftmargin=*,nosep]
\item An LLM-powered visual analytics framework for text-driven multi-criteria decision making. The system uses LLMs as a bridge to transform unstructured review text into a structured AHP decision model, supports users in iteratively inspecting and refining the model through interactive visual analytics, and ultimately generates a decision report containing rankings, criteria scores, and original review evidence, enabling complete traceability from results to supporting evidence.
\item A goal-oriented adaptive criteria discovery mechanism. Through coverage gap detection and data-driven exploration, the system helps users discover and incorporate decision dimensions not covered by the initial model, allowing the evaluation framework to dynamically align with users' personalized needs.
\item Two complementary AHP consistency repair approaches: an automated matrix adaptation algorithm that incorporates user weight adjustment intent, automatically adjusting the pairwise comparison matrix to maintain consistency when users drag weights; and a visual diagnosis based on joint Multidimensional Scaling (MDS) projection, helping users locate specific sources of inconsistency and make targeted corrections.
\end{itemize}

\section{Related Work}
\label{sec:related}

We review related work from three perspectives: LLM-based multi-criteria decision methods, visual analytics for decision processes, and recommender systems with visual analytics.

\subsection{LLM-Based Multi-Criteria Decision Methods}

Multi-criteria decision methods such as AHP have long relied on domain experts to manually construct criteria hierarchies and pairwise comparison matrices. Recently, researchers have begun automating this process with LLMs, which can be categorized by the role of the LLM. The first category replaces human experts with LLM parametric knowledge: Svoboda et al.~\cite{svoboda2024ahp} use GPT-4 to simulate virtual experts generating pairwise comparison matrices, and Vahidnia et al.~\cite{vahidnia2025multiagent} employ multi-agent negotiation across LLM platforms for spatial decision weighting. The second category fine-tunes LLMs for specific domains: Park et al.~\cite{park2025ahp} fine-tune an LLM on agricultural management documents for AHP evaluation (achieving 0.99 correlation with expert assessments), and Wang et al.~\cite{wang2025oneforall} use LoRA fine-tuning to reach expert-level scoring. The third category takes text documents as direct input: Doc2AHP~\cite{wu2026doc2ahp} performs embedding-based clustering on documents and uses LLMs to generate criteria labels and pairwise comparison matrices, achieving end-to-end conversion from text to a complete AHP model; Lu et al.~\cite{lu2024ahpllm} have the LLM simultaneously generate criteria and perform pairwise comparison scoring on open-ended responses.

These works demonstrate an evolution from expert-knowledge-driven to document-data-driven approaches, but they share a common focus on \textbf{automating the construction of decision models}---implicitly assuming the LLM-generated model is good enough for direct use. However, our quantitative experiment (\S\ref{sec:eval:quantitative}) shows that only 57.3\% of LLM-generated matrices satisfy consistency requirements, and in our case studies the initial model's coverage rate is below 60\%. Thus, \textbf{model generation is a starting point, not an endpoint}---we shift the emphasis to human-in-the-loop iterative refinement, supporting users in discovering coverage gaps, pruning hallucinated criteria, and progressively completing the model through visual analytics.

\subsection{Visual Analytics for Decision Processes}

Decision tasks have long held an implicit status in visualization research; Dimara \& Stasko~\cite{dimara2022decision} systematically show that most VA tools support decision processes limited to sorting and filtering predefined attributes rather than helping users construct evaluation frameworks. Multi-criteria decision making (MCDM) visualization has a long history. GAIA~\cite{mareschal2009gaia} uses principal component analysis (PCA) to project criteria and alternatives into a 2D space showing conflicts and synergies among criteria; Ishizaka et al.~\cite{ishizaka2016gaia} combine it with AHP for group preference evolution visualization. Asahi et al.~\cite{asahi1995treemap} use treemaps to display AHP decision trees with weight sensitivity analysis. Susmaga et al.~\cite{susmaga2024topsis} propose the WMSD space to make TOPSIS ranking processes explainable, and Sun et al.~\cite{sun2023fuzzyahp} combine intuitionistic fuzzy AHP with UMAP to display alternative distributions.

In multi-attribute ranking visualization, LineUp~\cite{gratzl2013lineup} supports interactive exploration of weighted rankings through adjustable-width stacked bars, ValueCharts~\cite{carenini2004valuecharts} uses tabular visualization for hierarchical objectives' weights and scores, WeightLifter~\cite{pajer2017weightlifter} visualizes the sensitivity of rankings to weight changes from a global weight-space perspective, and RankASco~\cite{schmid2022rankasco} maps raw data to user-preference-aligned scores via nonlinear scoring functions.

All these works operate on \textbf{predefined structured dimensions}. Gleicher et al.~\cite{gleicher2011comparison} systematize the design space of comparative visualization through juxtaposition, superposition, and explicit encoding strategies; our Scorecard View follows the juxtaposition strategy for multi-alternative comparison. However, when decision information resides in unstructured text, the dimensions themselves must be discovered and constructed---an upstream task not yet addressed by existing work. \sys{} incorporates dimension discovery into the visual analytics workflow, making it an integral part of decision exploration.

\subsection{Recommender Systems and Visual Analytics}

Recommender systems also involve user preference modeling and weight adjustment. Podium~\cite{wall2018podium} converts users' drag-to-rank operations into preference constraints and infers implicit weights via Ranking SVM; Gaggle~\cite{chang2020gaggle} uses Bayesian optimization to translate user interactions into model parameter feedback; PeerChooser~\cite{odonovan2008peerchooser} maps collaborative filtering similarity to node distances in a visual interface. For text integration, aspect-based opinion mining~\cite{zhang2023absa} systems extract multi-dimensional features from reviews for weighted ranking, and VisIRR~\cite{choo2018visirr} combines non-negative matrix factorization with PageRank for dynamic weighted recommendation.

The key difference lies in the user task. Recommender systems aim to \textbf{predict preferences and automatically generate recommendation lists}, with user intervention mainly correcting algorithmic bias. \sys{} targets \textbf{deliberative decision making}---users must actively understand each alternative's strengths and weaknesses, weigh trade-offs, and take responsibility for the final choice; weights are explicit components of the decision model rather than internal algorithm parameters.

In the broader text visual analytics field~\cite{liu2019bridging}, OpinionSeer~\cite{wu2010opinionseer} supports hotel review exploration but relies on predefined dimensions, while ConceptVector~\cite{park2018conceptvector} and El-Assady et al.~\cite{elassady2018progressive} respectively support interactive word embedding exploration and progressive topic modeling for users to incrementally construct and adjust semantic dimensions during analysis. Recently, LLM integration with visual analytics has intensified~\cite{endert2017integrating,shen2023nlvis,zhao2025leva}; \sys{} continues this trend, combining LLMs' text understanding with a structured decision framework, focusing on dynamically discovering evaluation dimensions from review text.

\section{Background and Design Requirements}
\label{sec:requirements}

\subsection{Background}
\label{sec:req:background}

We conducted multiple semi-structured interviews with three domain experts to understand practical needs and continuously collected feedback during prototype iterations: E1 is a professor in a school of management, specializing in multi-criteria decision methodology; E2 is the VP of Procurement \& Operations at a technology company, with long-term responsibility for office equipment procurement and supplier evaluation; E3 is a professor in a school of computer science, specializing in visualization and human-computer interaction.

E2 described a typical review-based decision scenario in the enterprise: procurement teams regularly purchase equipment for remote workers, and administrative teams select restaurants and hotels for corporate events. During the initial screening stage, procurement officers browse large numbers of reviews on e-commerce platforms and manually compile comparison tables for preliminary shortlisting. E2 identified three problems: different officers focus on different dimensions, leading to inconsistent results; reports must justify the selection to management but the tables often contain conclusions without supporting evidence; the team tried using ChatGPT to directly analyze reviews, but the generated reports were ``too generic---reasonable price, good quality---while the details I truly care about were completely missing.''

E1 summarized these problems as \textbf{incomplete dimensions} and \textbf{opaque processes}: when LLMs process large volumes of reviews, high-frequency topics mask low-frequency but important long-tail details, and even with carefully crafted prompt engineering, auto-generated models remain closed results that users cannot inspect. E2 added that criteria selection and weight assignment depend on the specific decision context---for the same batch of printers, ``color accuracy'' is key when purchasing for a design team, while ``duplex printing speed'' matters more for a finance department. E1 further noted that AHP requires users to perform pairwise comparisons among criteria and maintain consistency; when the number of criteria exceeds four, users find it difficult to locate inconsistent triads in the pairwise comparison matrix (i.e., transitivity violations where $a_{ij} \cdot a_{jk} \neq a_{ik}$); existing tools can flag which comparison values are inconsistent and suggest optimal corrections, but tracing the root cause of inconsistency still requires users to manually analyze the matrix. E3 noted from a visual analytics perspective that this workflow is essentially an iterative cognitive process of discovering evaluation dimensions from text, and what is lacking is effective tool support.

\subsection{Design Requirements}
\label{sec:req:requirements}

Based on the interviews and iterative design process, we derived the following design requirements:

\textbf{R1: Visualize criteria coverage and gaps.} The visualization should juxtapose the current criteria framework's coverage with the actual distribution of review data, enabling users to identify which review regions are not covered by any criterion. Uncovered regions should support further exploration through cues such as semantic clusters and keywords, allowing users to assess whether they contain long-tail dimensions worth incorporating.

\textbf{R2: Support discovering new criteria from uncovered reviews.} The visualization should provide data-driven exploration means to help users identify valuable subsets among uncovered reviews. The system should assist users in refining these subsets into new evaluation dimensions and integrating them into the existing criteria structure.

\textbf{R3: Display evidence quality within each criterion.} For each criterion, the visualization should present the quantity and quality distribution of supporting review evidence, as well as cross-alternative evidence volume differences. When an alternative lacks evidence under a criterion, the visualization should clearly indicate this uncertainty.

\textbf{R4: Provide iteration progress and model sufficiency feedback.} As users refine the criteria framework through multiple rounds, the visualization should provide clear progress indicators at each round---review utilization, ranking confidence, alternative discriminability---helping users decide whether further adjustment is needed.

\textbf{R5: Support interactive weight adjustment with consistency guarantees.} Users should be able to intuitively adjust relative weights among criteria to express personalized preferences, while the system provides real-time feedback on how weight adjustments affect rankings. The system should maintain AHP consistency while providing intuitive consistency diagnosis to help users understand the sources of inconsistency and participate in the correction process, rather than relying solely on black-box algorithms.

\textbf{R6: Support complete decision traceability and report generation.} Users should be able to start from an alternative's ranking result and drill down through a unified visual context: from the total score to the score composition across criteria, then to the specific review content supporting each score, understanding how the ranking is composed from evidence across dimensions. The system should also support exporting decision results as a structured report containing rankings, weights, scores, and key evidence summaries.

\subsection{System Overview}
\label{sec:req:overview}

\begin{figure}[t]
  \centering
  \includegraphics[width=0.9\linewidth]{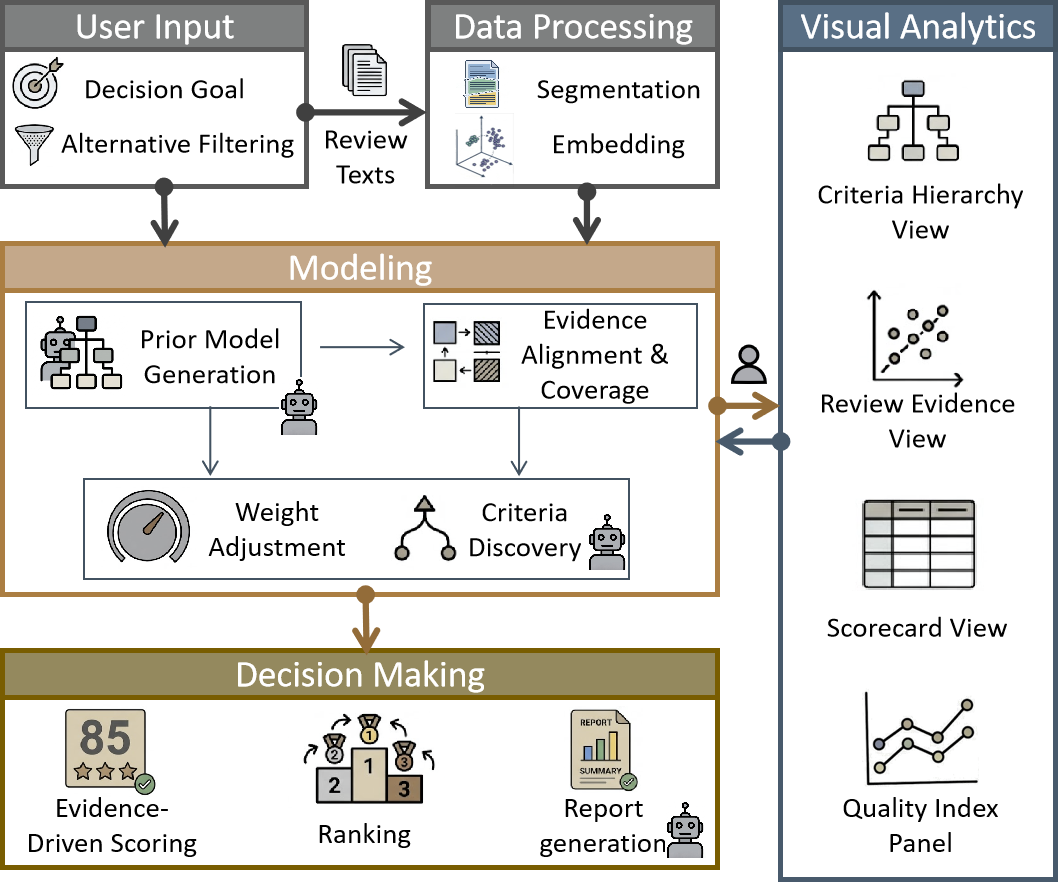}
  \caption{\sys{} system architecture. After the user specifies a decision goal, the data processing stage segments and encodes review text into a semantic space; the modeling stage generates a prior model via an LLM, performs evidence alignment, and enters an iterative loop of criteria discovery and weight adjustment; the decision output stage performs scoring, ranking, and report generation. Four visual analytics views on the right span the entire workflow, supporting exploration, diagnosis, and verification at each stage.}
  \label{fig:overview}
\end{figure}

Based on these design requirements, we designed the \sys{} system, whose overall architecture is shown in Fig.~\ref{fig:overview}. The workflow comprises four stages, with visual analytics views on the right providing interactive support throughout.

\textbf{User input and data processing.} The user specifies a decision goal and filters alternatives. The system segments review text into semantically focused snippets via sliding windows and encodes them into a shared semantic space using a pretrained embedding model (BGE-m3).

\textbf{Modeling iteration.} The system first uses an LLM to generate a prior model containing a criteria hierarchy and pairwise comparison matrices (\S\ref{sec:method}), then performs evidence alignment and coverage analysis. Users inspect coverage through the visual analytics interface and iterate between criteria discovery and weight adjustment: discovering new criteria expands the evaluation framework, while adjusting weights expresses personalized preferences---both jointly drive progressive model refinement.

\textbf{Decision output.} Once the model stabilizes, the system performs evidence-driven alternative scoring and ranking, and generates a decision report containing rankings, weight structures, and key evidence summaries.

\textbf{Visual analytics views} (\S\ref{sec:vis}) span the entire workflow: Criteria Hierarchy View supports criteria structure editing and consistency diagnosis, Review Evidence View supports coverage gap identification and criteria discovery, Scorecard View presents evidence chains and score structures, and Quality Index Panel monitors iteration progress.

\section{Method}
\label{sec:method}

This section describes the decision modeling methods of \sys{}. We draw on the analogy of Bayesian structure learning to organize the computational components for building personalized criteria frameworks from review text.

\subsection{Problem Formulation}
\label{sec:method:formulation}

Let the set of candidate alternatives be $A = \{a_1, \ldots, a_m\}$, each associated with a set of user reviews. Reviews are segmented into evidence snippets $D = \{d_1, \ldots, d_n\}$ via a sliding window, where each snippet $d_i$ carries its associated alternative $a(d_i) \in A$, rating $r(d_i)$, and sentiment score $s(d_i)$.

The system's goal is to construct an AHP decision model $\mathcal{M} = (\mathcal{G}, \theta, \phi)$, where $\mathcal{G}$ is the criteria hierarchy (a tree), $\theta$ is the criteria weight vector ($\sum_l \theta_l = 1$), and $\phi: D \to \mathcal{L}(\mathcal{G}) \cup \{\bot\}$ maps snippets to leaf criteria ($\bot$ denotes uncovered). Conceptually, a good model should satisfy two objectives simultaneously: covering as much evidence as possible (corresponding to the likelihood term in a Bayesian framework) and maintaining consistency with domain prior knowledge (corresponding to the prior term). This objective can be framed as the following Bayesian posterior:
\begin{equation}\label{eq:posterior}
  P(\mathcal{G}, \theta \mid D, K)
  \;\propto\;
  \underbrace{P(D \mid \mathcal{G}, \theta)}_{\text{evidence coverage}}
  \;\cdot\;
  \underbrace{P(\mathcal{G}, \theta \mid K)}_{\text{prior consistency}}
\end{equation}
where $K$ is the domain knowledge encoded in the LLM. Exact posterior inference over the combinatorial space of tree topologies is NP-hard; more importantly, the optimal criteria framework inherently depends on the user's specific decision context and personalized preferences, and cannot be determined by algorithms alone. \sys{} operationalizes this framework as a human-in-the-loop iterative process: the system identifies coverage gaps and presents them to users, who decide whether to incorporate new criteria (improving coverage); the system then reintegrates weights under AHP consistency constraints (maintaining prior consistency). Each iteration observably increases coverage while consistency constraints ensure mathematical soundness.

Given a model $\mathcal{M}^* = (\mathcal{G}^*, \theta^*, \phi^*)$ the user is satisfied with, the final ranking is obtained by weighted aggregation:
\begin{equation}\label{eq:score}
  \mathrm{score}(a_j) = \sum_{l \in \mathcal{L}(\mathcal{G}^*)} \theta^*_l \cdot f(a_j, l)
\end{equation}
where $\mathcal{L}(\mathcal{G}^*)$ is the set of leaf criteria and $f(a_j, l)$ is the score of alternative $a_j$ on leaf criterion $l$ (\S\ref{sec:method:scoring}). This separation makes \textbf{evidence} (criterion-level scores $f$) orthogonal to \textbf{preferences} (weights $\theta$)---the same scores produce different rankings under different weights, so users can adjust weights without re-evaluating evidence, reducing iteration cost. The following subsections describe each computational component.

\subsection{Prior Model Generation}
\label{sec:method:prior}

Given the user's natural-language decision objective $q$ (e.g., ``select all-in-one printers for a remote work team''), the system uses the LLM to generate an initial model $\mathcal{M}_0 = (\mathcal{G}_0, \theta_0)$.

The LLM generates an initial criteria tree $\mathcal{G}_0$ from $q$ via chain-of-thought reasoning~\cite{wei2022cot} (4--6 top-level criteria, each with 2--4 leaf criteria), with prompts requiring the LLM to justify mutual exclusivity among criteria (MECE principle) and measurability of leaf criteria step by step. For each group of sibling criteria, the LLM performs pairwise comparisons on the Saaty 1--9 scale, generating a reciprocal matrix $A$ ($A_{ij} \cdot A_{ji} = 1$). This approach uses the LLM as a pairwise judge, whose reliability has been systematically studied by Zheng et al.~\cite{zheng2023llmjudge}---they find that strong LLMs' pairwise preferences show high agreement with human evaluations but also exhibit position and verbosity biases, which is precisely why we introduce consistency repair (\S\ref{sec:method:weight}) and human review. Initial weights $\theta_0$ are computed via the logarithmic least squares method (LLSM)~\cite{crawford1985llsm}, and the consistency ratio ($\mathrm{CR} < 0.1$) is verified. Global weights are obtained by multiplying local weights along root-to-leaf paths.

\subsection{Evidence Embedding and Semantic Clustering}
\label{sec:method:embedding}

The system filters relevant alternatives and their reviews from the dataset based on the decision objective, and segments reviews into snippets via a sliding window (window size $w=2$, minimum length threshold $l_{\min}$). Snippets are encoded into a high-dimensional semantic space using a pretrained dense retriever (BAAI/bge-m3~\cite{chen2024bge}, 568M parameters), then projected to 2D coordinates via UMAP~\cite{mcinnes2018umap} for visual exploration.

HDBSCAN~\cite{campello2013hdbscan} is applied on the 2D projection to perform density-based clustering on all snippets. HDBSCAN automatically identifies clusters of varying densities without requiring a preset number of clusters; snippets not absorbed by any cluster are marked as noise points ($\mathrm{label} = -1$). TF-IDF~\cite{sparckjones1972idf} keywords are extracted for each cluster as interpretable labels. Additionally, VADER~\cite{hutto2014vader} sentiment analysis is applied to all snippets to compute sentiment scores (compound score $\in [-1, 1]$).

\subsection{Evidence Coverage Analysis}
\label{sec:method:coverage}

Review regions not covered by any criterion represent weak spots in the model's explanatory power. This section converts this deficiency into actionable visual signals. Performing coverage analysis in the embedding space rather than via topic models is a key design decision: topic models (e.g., LDA) produce fuzzy topic boundaries at fixed granularity, making it hard to distinguish fine-grained differences such as ``printhead clogging'' versus ``ink drying''; the embedding space preserves semantic proximity, so density clusters and gaps directly correspond to the coverage--blind-spot structure of the criteria framework.

For each semantic cluster, the system submits its keywords and representative snippets along with the current list of leaf criteria to the LLM, which judges the cluster's relevance to each criterion ($0$=unrelated, $1$=partially related, $2$=directly related). The criterion with the highest relevance above a threshold becomes the cluster's assigned criterion; clusters not sufficiently explained by any criterion are marked as uncovered. Coverage $\mathrm{coverage}(\mathcal{M}) = 1 - |D_{\bot}| / |D|$ measures the criteria framework's overall explanatory power over the corpus, where $D_{\bot}$ is the set of snippets belonging to uncovered clusters and noise points.

The system also computes novelty scores for noise points to quantify their potential for harboring new dimensions. For each noise point $d_i$, the maximum cosine similarity with all cluster centroids in the high-dimensional embedding space is computed as $\mathrm{max\_sim}_i$, and novelty is defined as:
\begin{equation}\label{eq:novelty}
  \mathrm{novelty}(d_i) = \mathrm{normalize}\bigl(1 - \mathrm{max\_sim}_i\bigr)
\end{equation}
Normalization uses 5th--95th percentile clipping to suppress the influence of outliers. High-novelty noise points are dissimilar to all known semantic clusters and are the regions most likely to contain new evaluation dimensions.

\subsection{Criteria Discovery and Structure Update}
\label{sec:method:discovery}

After users identify relevant uncovered regions, the system assists in converting them into new criteria and integrating them into the model.

Criteria discovery proceeds in two steps. The system submits the text, sentiment distribution, rating distribution, and current AHP tree structure of selected snippets to the LLM, which identifies 3--6 candidate topics (with names, descriptions, keywords, and estimated proportions). The LLM then proposes 2--3 new criteria based on these candidates, specifying the best parent node for attachment; sub-criteria are generated when necessary to align with existing branch depth. These numeric ranges are empirically set; in case studies and the user study, users actually adopted approximately 1--2 criteria per round. After user confirmation, new criteria are inserted into $\mathcal{G}$, completing the structure update $\mathcal{G}_t \to \mathcal{G}_{t+1}$.

After a new criterion is added, the LLM performs pairwise comparisons between it and each sibling, expanding the original $n \times n$ matrix to $(n{+}1) \times (n{+}1)$. Weights are recomputed via LLSM and consistency is verified.

\subsection{Weight Adjustment and Consistency Guarantee}
\label{sec:method:weight}

After exploring evidence, users often form preferences that differ from the LLM's initial weights. The core challenge of weight adjustment is that interactive fine-tuning must respect user intent while maintaining AHP's mathematical consistency ($\mathrm{CR} \leq 0.1$).

We formalize this as a constrained optimization problem. Given the user's adjusted target weights $\theta^{\mathrm{target}}$, the system constructs an ideal comparison matrix $A^{\mathrm{ideal}}_{ij} = \theta^{\mathrm{target}}_i / \theta^{\mathrm{target}}_j$ and maps each element to the Saaty 1--9 scale (maintaining reciprocity $A_{ji} = 1/A_{ij}$). Since discretization may introduce inconsistency, the system solves:
\begin{equation}\label{eq:weight_opt}
  A^* = \arg\min_{A \in \mathcal{S}} \sum_{i<j} \bigl(\ln A_{ij} - \ln A^{\mathrm{ideal}}_{ij}\bigr)^2, \quad \text{s.t. } \mathrm{CR}(A^*) \leq 0.1
\end{equation}
where $\mathcal{S}$ is the space of reciprocal Saaty-scale matrices. The objective minimizes deviation in log-space, ensuring repaired weights stay as close as possible to the user's intent. We choose to solve on the discrete Saaty scale rather than in continuous space---a key design decision for human-computer interaction: continuous-space repair methods (e.g., geometric mean adjustment~\cite{crawford1985llsm}) produce comparison values that are unintuitive decimals (e.g., $2.37$ or $1/4.18$) that users cannot understand or verify; Saaty-scale integers and their reciprocals (e.g., ``A is 3 times more important than B'') carry clear semantics, enabling users to understand and inspect each comparison value's reasonableness. The solver operates directly on the discrete feasible set $\mathcal{S}$: elements are sorted by their sensitivity to CR, and the most impactful upper-triangular elements are adjusted first, iterating over the 17-level Saaty values ($\{1/9,\ldots,1,\ldots,9\}$) until the constraint is satisfied. Since the search space is combinatorial, local search may theoretically stop at a local optimum, but at practical scales (each sibling group has $n \leq 6$), all 117 matrices in our quantitative experiment (\S\ref{sec:eval:quantitative}) were successfully repaired. The repaired matrix yields final weights via LLSM.

\sys{} offers two complementary weight adjustment modes: (1) users drag weight bars to set target weights and the system automatically executes the above optimization, suitable for quick adjustments; (2) users diagnose inconsistency sources in the joint MDS space and modify pairwise comparison values (\S\ref{sec:vis:tree}), suitable for in-depth understanding of weight structure. Both modes ensure mathematical soundness of final weights through Eq.~\eqref{eq:weight_opt}.

\subsection{Evidence-Driven Alternative Scoring}
\label{sec:method:scoring}

Given the user's iteratively refined model $(\mathcal{G}^*, \theta^*)$ and mapping $\phi$, the system computes each alternative's score on every criterion. At the prior model generation stage, all criteria weights are equally distributed ($\theta_l = 1/|\mathcal{L}(\mathcal{G})|$) and alternative scores are simply the mean of all their review ratings---this initial state provides an unbiased starting point, with weights and coverage progressively differentiating as users iterate.

For each (alternative $a_j$, leaf criterion $l$) pair, all snippets mapped to that criterion are collected as $S_{jl} = \{d_i : \phi(d_i) = l,\; a(d_i) = a_j\}$, and the mean of their original review ratings serves as the score:
\begin{equation}\label{eq:criterion_score}
  f(a_j, l) = \frac{1}{|S_{jl}|} \sum_{d_i \in S_{jl}} r(d_i)
\end{equation}
where $r(d_i)$ is the original rating of the review containing the snippet (e.g., Amazon's 1--5 stars). The global score is computed by weighted summation via Eq.~\eqref{eq:score}, and alternatives are ranked in descending order.

Note that $r(d_i)$ is the reviewer's overall product rating rather than a criterion-specific rating. This simplification is acceptable in practice: after HDBSCAN semantic clustering, snippets within the same cluster are topically coherent with small rating variance; moreover, the beeswarm plots in the Scorecard View (\S\ref{sec:vis:evidence}) visualize each snippet's rating distribution, letting users assess the reliability of aggregated scores. Future work could explore having the LLM generate criterion-level ratings for each snippet to improve precision.

\section{Visualization Design}
\label{sec:vis}

The \sys{} interface consists of four coordinated regions (Fig.~\ref{fig:teaser}): the left side houses the Control Panel~(A) and Quality Index Panel~(B) for parameter settings and quality monitoring; the upper center contains the Review Evidence View~(D) and AHP Hierarchy View~(C); the lower center contains the Scorecard View~(E); the right side contains the LLM Analysis Panel~(D$_2$) and Decision Report~(A$_1$). All views share a unified color scheme---each top-level criterion is assigned a primary color, with its leaf criteria taking lighter/darker variants---ensuring cross-view visual consistency. The following subsections describe each component.

\subsection{Review Evidence View}
\label{sec:vis:scatter}

The Review Evidence View (Fig.~\ref{fig:teaser}D) projects all review snippets into a 2D semantic space and serves as the primary analysis canvas for coverage gap identification (\textbf{R1}) and criteria discovery (\textbf{R2}). The design draws on embedding space comparison visualization~\cite{boggust2022embedding}, which reveals semantic relationships through global structure and local neighborhoods, adapted here for coverage gap analysis.

The scatter plot supports two complementary modes, each serving a distinct cognitive activity in decision analysis. \emph{Coverage mode} (Fig.~\ref{fig:teaser}D$_1$) answers ``what does the current framework already explain'': covered snippets are colored by assigned criterion while uncovered noise points are rendered in gray; selecting a top-level criterion displays its snippets at full opacity with others dimmed. This creates a figure-ground contrast between coverage and gaps, enabling users to grasp the overall picture without checking individual snippets. \emph{Explore mode} (Fig.~\ref{fig:teaser}D) answers ``what might be missing that deserves attention'': the system overlays KDE-based contour layers to reveal spatial aggregation trends among noise points, with TF-IDF keywords (preferring bigrams for both readability and descriptive power) placed at density peaks. Separating these activities into distinct modes rather than overlaying them is motivated by pilot testing, which revealed that simultaneously displaying criterion coloring and density contours caused severe visual interference---users could neither effectively assess coverage quality nor focus on exploring blind spots. Users select snippets via lasso as input for criteria discovery (\S\ref{sec:method:discovery}).

\subsection{AHP Hierarchy View}
\label{sec:vis:tree}

The AHP Hierarchy View (Fig.~\ref{fig:teaser}C) presents the decision model's criteria hierarchy, weight allocation, and consistency status as an interactive tree, supporting criteria structure editing (\textbf{R2}) and weight adjustment (\textbf{R5}).

The tree expands from the top goal node down to top-level and leaf criteria. Edge stroke width encodes global weight (product of local weights along the root-to-leaf path), letting users perceive each leaf criterion's actual contribution to the final ranking from visual thickness without mental multiplication. Weight fill bars within nodes encode local weight (share among siblings), complementing the global view---the former answers ``how important is this criterion overall'' while the latter answers ``what proportion does it occupy among its siblings.'' Below each sibling group, the consistency ratio CR is annotated: green when $\mathrm{CR} \leq 0.1$, red when above threshold to alert the user. Users can drag weight bars to adjust weights; the system displays before/after comparison marks in real time and triggers consistency repair (\S\ref{sec:method:weight}) on release. Clicking a top-level criterion highlights its snippets in the scatter plot; double-clicking adds it to the LLM analysis context. Newly grafted criteria are marked with dashed borders until confirmed.

\begin{figure*}[t]
  \centering
  \includegraphics[width=\linewidth]{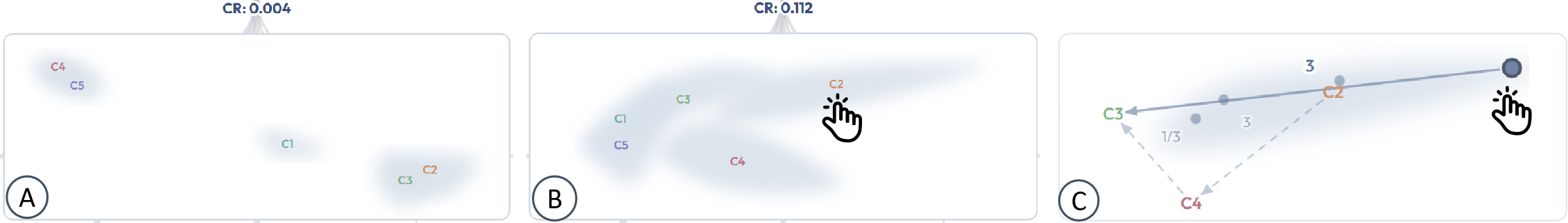}
  \caption{Consistency Diagnosis View. Each criterion has $n$ projection points (corresponding to $n$ perspectives); their spread encodes consistency: the smaller the CR, the more consistent the distance matrices across perspectives and the tighter the projection points cluster; the larger the CR, the more the points spread apart, with the most spread-out criterion being the primary source of inconsistency. (A)~A consistent matrix ($\mathrm{CR}=0.004$): projection points for each criterion cluster tightly. (B)~An inconsistent matrix ($\mathrm{CR}=0.112$): dashed ellipses mark the most spread-out criteria. (C)~Diagnostic details: solid arrows annotate direct comparison values, dashed arrows show indirect derivation paths, and the discrepancy between them indicates transitivity violations.}
  \label{fig:mds}
\end{figure*}

Clicking the CR label expands the Consistency Diagnosis View (Fig.~\ref{fig:teaser}C$_1$), mapping consistency status to intuitive spatial patterns (\textbf{R5}). The core challenge of AHP consistency diagnosis is that CR only provides a global score---users cannot determine which specific comparison values produce contradictions. Becker et al.~\cite{becker2024graphical} propose arranging criteria on a one-dimensional number line to visualize comparison relationships, but a 1D arrangement cannot simultaneously show conflicts across multiple perspectives. We extend this to a two-dimensional joint MDS space. The core idea treats each column of the comparison matrix as a ``perspective'' $j$---column $j$ reflects the distance structure of ``comparing all criteria with criterion $j$ as the reference.'' The log-distance between criteria within a perspective is $d_{ik}^{(j)} = |\ln a_{ij} - \ln a_{kj}|$, and all $n$ criteria across all perspectives are simultaneously embedded into a shared 2D space via joint MDS~\cite{borg2005mds}:
\begin{equation}\label{eq:mds}
  L = \underbrace{\sum_{j}\sum_{i<k}\bigl(\|p_i^{(j)} - p_k^{(j)}\| - d_{ik}^{(j)}\bigr)^2}_{\text{MDS stress}} + \lambda \underbrace{\sum_{i}\sum_{j<l}\|p_i^{(j)} - p_i^{(l)}\|^2}_{\text{cross-perspective alignment}}
\end{equation}
When the matrix is perfectly consistent ($a_{ij} \cdot a_{jk} = a_{ik}$), the distance matrices from all perspectives are identical and the $n$ projection points for each criterion coincide; inconsistency causes points to spread, with the degree of spread directly quantifying that criterion's inconsistency (Fig.~\ref{fig:mds}A--B).

Clicking a spread-out criterion expands diagnostic details (Fig.~\ref{fig:mds}C): solid arrows annotate direct comparison values $a_{ij}$, dashed arrows show indirect derivation paths $a_{ik} \cdot a_{kj}$, and the discrepancy is the degree of transitivity violation. Arrow directions are normalized: when a majority of values in a triad are fractions ($<1$), the system automatically flips arrow directions so that all arrows consistently point toward the ``more important'' side, avoiding directional confusion. When the user hovers over a Saaty scale value on a connection line, $+$/$-$ buttons appear on either side; each click steps one position along the 17-level Saaty scale ($1/9, 1/8, \ldots, 1, 2, \ldots, 9$). After each adjustment, the system refines the MDS projection via warm-start (150 iterations at a reduced learning rate) from the current layout. This design choice is critical: recomputing from scratch would cause layout jumps due to MDS's multiple-solution nature, causing users to lose their spatial frame of reference; warm-start ensures smooth layout evolution, letting users judge each adjustment's effect by observing how projection points converge---transforming consistency repair from ``adjust parameters and check numbers'' to the intuitive operation of ``adjust parameters and watch spatial changes.'' This targeted correction complements the automatic repair in \S\ref{sec:method:weight}. The alignment regularization coefficient $\lambda$ (Eq.~\eqref{eq:mds}) is set to 1.0; this value was stable across matrices of different sizes in our experiments.

\subsection{Scorecard View}
\label{sec:vis:evidence}

The Scorecard View (Fig.~\ref{fig:teaser}E) displays each alternative's ranking, criterion scores, and evidence distribution in a three-column side-by-side layout, serving as the primary view for inspecting scoring evidence (\textbf{R3}), monitoring iteration progress (\textbf{R4}), and tracing decision rationale (\textbf{R6}).

Each row corresponds to one alternative, and the three columns form a progressive information hierarchy from left to right: ranking result (overview) $\to$ score composition across criteria (pattern comparison) $\to$ evidence distribution supporting scores (verification). Aligning three layers horizontally within the same row maintains visual continuity, allowing users to grasp each alternative's complete information---from ranking to score structure to evidence distribution---at a glance, consistent with Gleicher et al.'s~\cite{gleicher2011comparison} finding that juxtaposition is best suited for attribute-by-attribute comparison.

The left column (E$_1$) shows ranking number, name, price, and base score; alternatives that have entered the top five receive persistent color identifiers for tracking. The middle column (E$_2$) is the criterion score matrix: each column corresponds to a leaf criterion, each row to an alternative, with cells encoding scores as stacked color blocks. Three column arrangement modes serve different analysis tasks: \emph{uniform mode}~(E$_2$) with equal-width columns for per-criterion comparison across alternatives; \emph{stacked mode}~(E$_3$) with column widths scaled by weight, borrowing LineUp's~\cite{gratzl2013lineup} strategy of using visual area to encode weighted contribution so that ``visually wider criteria indeed matter more'' and users can judge ranking causes from total bar width; and \emph{detail mode} with fixed widths and horizontal scrolling for full numerical values. When an alternative lacks evidence under a criterion, the cell is marked with a hatched fill pattern indicating uncertainty. Users can drag column headers to adjust weights, synchronized with the AHP tree's weight editing.

The right column (E$_4$, E$_5$) uses beeswarm plots with kernel density overlays to display review snippet distributions for each alternative. We chose beeswarm plots over box plots or violin plots for two reasons: first, evidence traceability (\textbf{R6}) requires users to drill down to individual reviews---box plots discard individual data, violin plots show density but do not support interactive selection, while beeswarm plots preserve each data point's independent selectability; second, snippets in the beeswarm share the same visual encoding (shape, color, selection state) as in the Review Evidence View scatter plot, so users switching between views need not relearn the mapping, reducing the cognitive cost of cross-view association. Snippets use force-directed layout~\cite{bostock2011d3} to avoid overlap, with spatially adjacent snippets merged to reduce clutter; merged markers encode aggregate count by size. Snippets are colored by assigned criterion, and the horizontal axis supports switching among rating, sentiment, and recency dimensions; column headers show miniature histograms previewing the global distribution.

\subsection{Control Panel and Quality Index Panel}
\label{sec:vis:control}

The left panels (Fig.~\ref{fig:teaser}A, B) provide parameter control and quality monitoring (\textbf{R4, R5}). The Control Panel~(A) contains dataset selection and price/rating filters. The Quality Index Panel~(B) has two components: a score evolution ribbon chart tracking score changes and rank shifts of the top five alternatives across iteration steps, and a maximum CR line chart showing the consistency ratio after each adjustment (with $\mathrm{CR} = 0.1$ as the threshold), helping users monitor the model's mathematical soundness.

\subsection{LLM Analysis Panel}
\label{sec:vis:chat}

When users lasso-select snippets in the scatter plot, the system passes the selected snippets' original text, statistical summary (count, sentiment distribution, rating distribution, number of affected alternatives), and the current AHP criteria tree structure to the LLM. The LLM generates structured analysis results, which the LLM Analysis Panel (Fig.~\ref{fig:teaser}D$_2$) presents in a card-based layout: a thematic summary of the selected region, representative review quotes, and candidate criteria suggestions (with confidence scores and suggested attachment positions in the criteria tree). Candidate criteria are checked against existing criteria to avoid duplication. User confirmation triggers the criteria grafting workflow (\S\ref{sec:method:discovery}). Conversation history is saved as a session list for retrospective comparison.

\subsection{Decision Report}
\label{sec:vis:report}

The Decision Report (Fig.~\ref{fig:teaser}A$_1$) opens as a modal drawer, generating an exportable structured decision document (\textbf{R6}) containing an executive summary, ranking table (top-ten alternatives' criterion scores), weight treemap (with CR annotations), radar chart (multi-dimensional comparison of the top two), and evidence summary (representative positive/negative review quotes).

\subsection{Typical Workflow}
\label{sec:vis:workflow}

A typical usage session follows an ``explore--adjust--verify'' iterative pattern. Users first identify uncovered regions in the Review Evidence View's explore mode, lasso-select snippets of interest, and use LLM analysis to confirm new criteria for grafting into the AHP tree. They then adjust weights in the AHP Hierarchy View to express personalized preferences, using the Consistency Diagnosis View to locate and correct inconsistency sources when needed. After each round, users inspect score structures and evidence distributions in the Scorecard View and monitor coverage and ranking stability via the Quality Index Panel. When rankings converge and coverage reaches a satisfactory level, users generate the Decision Report to finalize the decision.

\section{Evaluation}
\label{sec:eval}

We evaluate \sys{} from three perspectives, corresponding to the VDAR (Visual Data Analysis and Reasoning) and UWP (Understanding Work Practices) scenarios identified by Lam et al.~\cite{lam2011empirical}: two case studies in different domains demonstrate the system's practical utility, a user study validates usability for target users, and a quantitative experiment verifies the consistency repair mechanism. In-depth feedback from the collaborating experts (E1--E3) who participated in the design process is discussed in \S\ref{sec:discussion}. In all evaluations, the system's LLM components (prior model generation, cluster-criteria mapping, criteria discovery) use GPT-5.2, and text embedding uses BGE-m3 (\S\ref{sec:method:embedding}).

\subsection{Datasets}
\label{sec:eval:data}

We use two review datasets from different sources to verify cross-domain applicability.

\textbf{Amazon Office Products.} From the ``5-core'' subset of Amazon Review Data~\cite{ni2019justifying}, filtering for printer products. Contains 4,627 English reviews for 29 printers. This dataset is used for Case Study 1, the user study, and the quantitative experiment.

\textbf{HotelRec.} From the HotelRec~\cite{antognini2020hotelrec} hotel review dataset, filtering for hotels in the Paris area. Contains 5,159 English reviews for 20 hotels. This dataset is used for Case Study 2 and the user study.

Both datasets underwent identical preprocessing: review text was segmented into semantic snippets via sliding windows, encoded into a high-dimensional embedding space via BGE-m3, then projected via UMAP and clustered via HDBSCAN to generate spatial coordinates and cluster labels for visualization.

\subsection{Case Study 1: Remote Office Printer Selection}
\label{sec:eval:case1}

\begin{figure}[t]
  \centering
  \includegraphics[width=\linewidth]{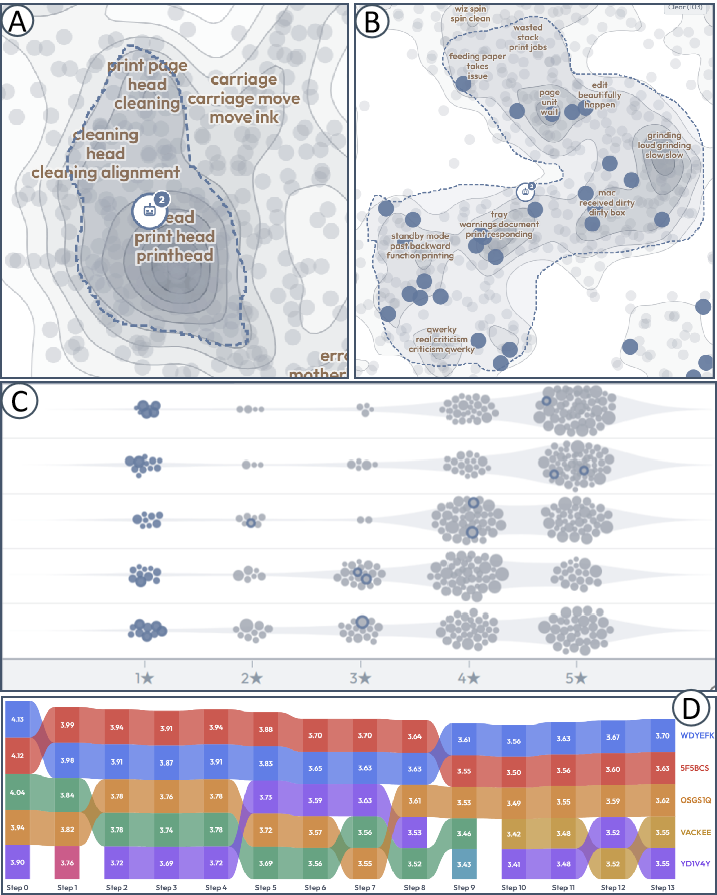}
  \caption{Case Study 1. (A)~Lasso selection over uncovered regions in explore mode; keywords reveal printhead-related topics. (B)~Another cluster points to mechanical jams and paper handling failures. (C)~Beeswarm plots show that top-5 products still have many 1-star reviews. (D)~Score evolution ribbon chart tracking ranking changes across 13 iterations; top 3 stabilize in later rounds.}
  \label{fig:case1}
\end{figure}

The first case targets a procurement officer selecting all-in-one printers for a remote work team, using the Amazon printer dataset (\S\ref{sec:eval:data}). After E2 entered the procurement requirements, the system generated an initial AHP model with 5 top-level criteria (Print Performance, Scanning Capability, Connectivity \& Compatibility, Operating Costs, Physical \& Practical Fit) and 15 leaf criteria. E2 used the price filter to narrow candidates to 14 printers.

At this point, the Review Evidence View showed a coverage rate of only 59.1\% (\textbf{R1}), with gray uncovered regions forming multiple clusters. The initial model already covered explicitly stated requirements---WiFi Stability, MacOS Support, Color Print Quality---but uncovered regions still accounted for over 40\%. E2 switched to explore mode and lasso-selected 158 review snippets in the densest region (Fig.~\ref{fig:case1}A), where contour lines and keyword labels (``print head,'' ``cleaning,'' ``carriage move'') revealed printhead-related topics (\textbf{R1}). The LLM Analysis Panel showed that most snippets in the selection discussed printhead clogging, failure, and cleaning costs---a dimension not explicitly mentioned in the decision objective but with decisive impact on user experience. E2 adopted the ``Printhead Reliability'' criterion and grafted it under Print Performance (\textbf{R2}). Rankings changed immediately: the former second-place product surpassed the first due to higher printhead reliability scores (Fig.~\ref{fig:case1}D, \textbf{R4}).

E2 then observed from the Scorecard View's beeswarm plots that the top-5 products still had a substantial number of 1-star reviews (Fig.~\ref{fig:case1}C, \textbf{R3}), and located these negative reviews clustering in the lower-right region of the Review Evidence View (Fig.~\ref{fig:case1}B). Zooming in revealed keyword labels ``grinding,'' ``feeding paper,'' ``spin clean,'' pointing to mechanical failures. The LLM analysis report indicated that 60\% of snippets highlighted frequent paper jams, paper wrinkling, and structural fragility, leading E2 to adopt a ``Paper Handling Reliability'' criterion (\textbf{R2}).

Through the same exploration workflow, E2 successively discovered and added Scanner Software Usability, Ink/Toner Operating Cost, and Initial Setup Complexity criteria. After 13 iterations, coverage increased from 59.1\% to 68.3\%, the top 3 rankings stabilized (Fig.~\ref{fig:case1}D, \textbf{R4}), and E2 stopped iterating and generated a decision report (\textbf{R6}).

\subsection{Case Study 2: Paris Business Trip Hotel Selection}
\label{sec:eval:case2}

\begin{figure}[t]
  \centering
  \includegraphics[width=\linewidth]{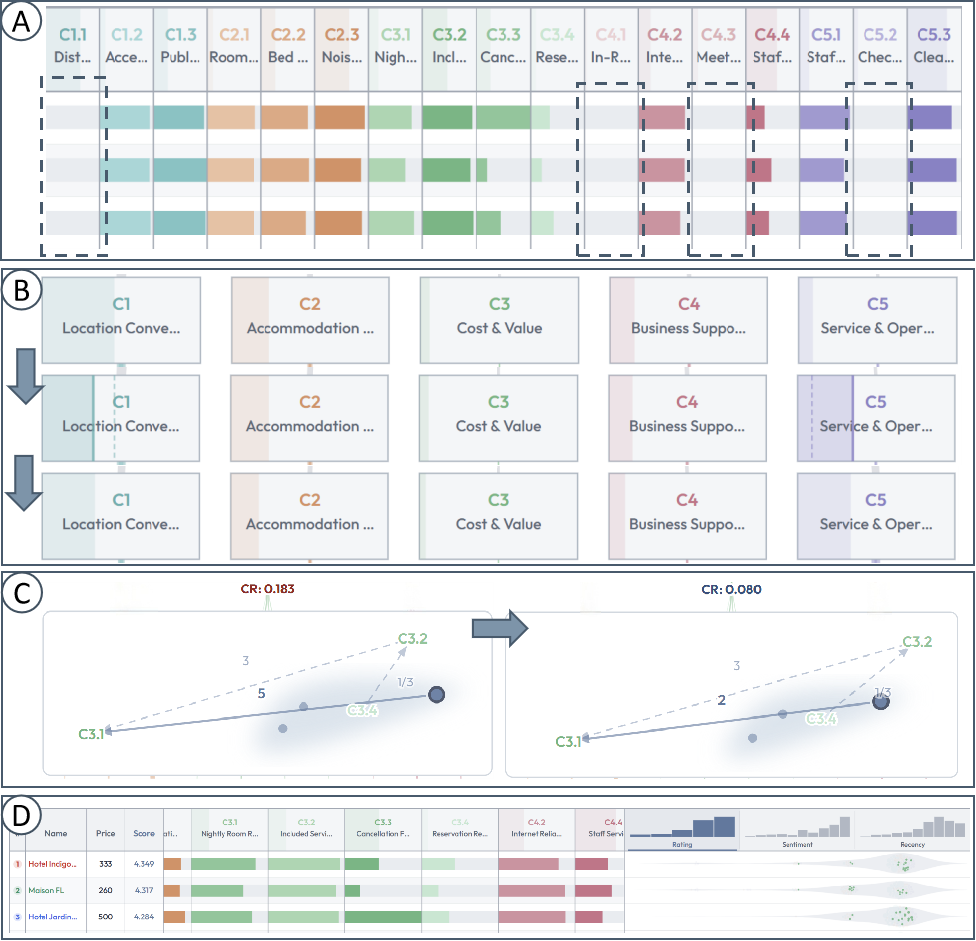}
  \caption{Case Study 2. (A)~Four zero-match criteria are deleted. (B)~Weight adjustment: reducing C1, increasing C5. (C)~Consistency repair: C3's CR drops from 0.183 to 0.080. (D)~Criterion scores and beeswarm plot comparison for the top-3 hotels.}
  \label{fig:case2}
\end{figure}

The second case targets an administrative officer selecting Paris hotels for an executive team business trip, using the HotelRec Paris hotel dataset (\S\ref{sec:eval:data}). E1 entered the following goal: ``Selecting hotels for a company executive team 3-day business trip to Paris. 4 travelers, good location preferred, quiet and comfortable, moderate budget.'' The system generated an initial AHP model with 5 top-level criteria: Location Convenience (C1), Accommodation Comfort (C2), Cost \& Value (C3), Business Support (C4), and Service \& Operations (C5). E1 filtered to 12 top-rated hotels.

E1 first noticed that the top-5 hotels' scores were extremely close (4.450--4.618), with beeswarm plots showing ratings almost entirely concentrated in the 4--5 star range (\textbf{R3}), making differentiation by mean alone difficult. E1 turned to exploring uncovered regions (\textbf{R1}), focusing on the few low-rated reviews and the most recent ones, successively adding Staff Service Quality, Reservation Reliability, Breakfast Experience Quality, and Walkable Local Amenities criteria (\textbf{R2}), raising coverage to 69.5\%.

Meanwhile, E1 observed in the Scorecard View that 4 criteria (Distance to Meetings, In-Room Workspace, Meeting Facilities, Check-in Efficiency) had zero review matches (Fig.~\ref{fig:case2}A, \textbf{R3}), judging them to be hypothetical dimensions inferred by the LLM from the ``business travel'' semantics---no actual reviews discussed these topics---and deleted them from the model. This bidirectional operation demonstrates the system's support for users to both supplement long-tail dimensions and prune evidence-lacking ``hallucinated criteria.''

E1 then adjusted weights: reducing C1 and increasing C5 to match core priorities (Fig.~\ref{fig:case2}B, \textbf{R5}). After adjustment, CR values for C3 and C4 exceeded the 0.1 threshold. E1 clicked the CR label for C3 to expand the Consistency Diagnosis View (Fig.~\ref{fig:case2}C) and immediately noticed that C3.4 Cancellation Flexibility's projection points were the most spread out---its four perspective points formed a scatter area far larger than other criteria, indicating the primary source of inconsistency. E1 clicked C3.4 to expand diagnostic details and found that the solid arrow (direct comparison) and dashed arrow (indirect path) pointing to C3.1 ran in opposite directions with vastly different lengths---the visual conflict was immediately apparent. E1 used the $+$/$-$ buttons on the connection line to step the value down; after each click, the MDS layout smoothly updated via warm-start and the projection points gradually converged. After three adjustments, CR dropped from 0.183 to 0.080; a similar visual diagnosis and repair on C4 reduced CR from 0.117 to 0.065 (\textbf{R5}).

After these adjustments, the score gap among the top-3 hotels widened from 0.002 to 0.032, significantly improving discriminability (\textbf{R4}). E1 compared the score structures of top 1 and top 2 in the Scorecard View (Fig.~\ref{fig:case2}D), finding the gap mainly from C3's Cancellation Flexibility and Reservation Reliability dimensions; beeswarm plots confirmed that top 1 had more 5-star reviews on these dimensions (\textbf{R6}). After verifying that review content matched scores, E1 generated a decision report (\textbf{R6}).

\subsection{User Study}
\label{sec:eval:userstudy}

To evaluate \sys{}'s usability for target users, we recruited 8 participants (P1--P8, 6 male, 2 female, ages 23--28). Participants were master's and doctoral students in computer science or management with everyday experience using online reviews for purchase decisions; some had basic knowledge of AHP or MCDM methods. Participants were randomly divided into two groups: P1--P4 used the HotelRec Paris hotel dataset, and P5--P8 used the Amazon office products dataset.

Each session lasted approximately 55 minutes. The researcher first spent 10 minutes introducing AHP basics (criteria hierarchy, weights, consistency) and demonstrating system operations. Participants then completed a 15-minute guided task to familiarize themselves with core interactions---including viewing the criteria tree, exploring uncovered regions in the scatter plot, lasso selection and criteria grafting, and weight dragging. This was followed by a 20-minute free exploration phase. We asked participants to adopt realistic decision scenarios: hotel group participants assumed they were booking a Paris hotel for a business trip, while office products group participants assumed they were procuring equipment for their team. Participants independently set decision objectives and completed the full workflow from criteria discovery to ranking generation, using a think-aloud protocol~\cite{ericsson1993protocol} throughout. Finally, participants completed a 12-item 5-point Likert scale questionnaire (1=strongly disagree, 5=strongly agree) and a 10-minute semi-structured interview. The questionnaire covered four dimensions: criteria discovery effectiveness (Q1--Q3), weight adjustment and consistency (Q4--Q6), evidence traceability and trust (Q7--Q9), and overall usability (Q10--Q12).

\begin{figure}[t]
  \centering
  \includegraphics[width=0.9\linewidth]{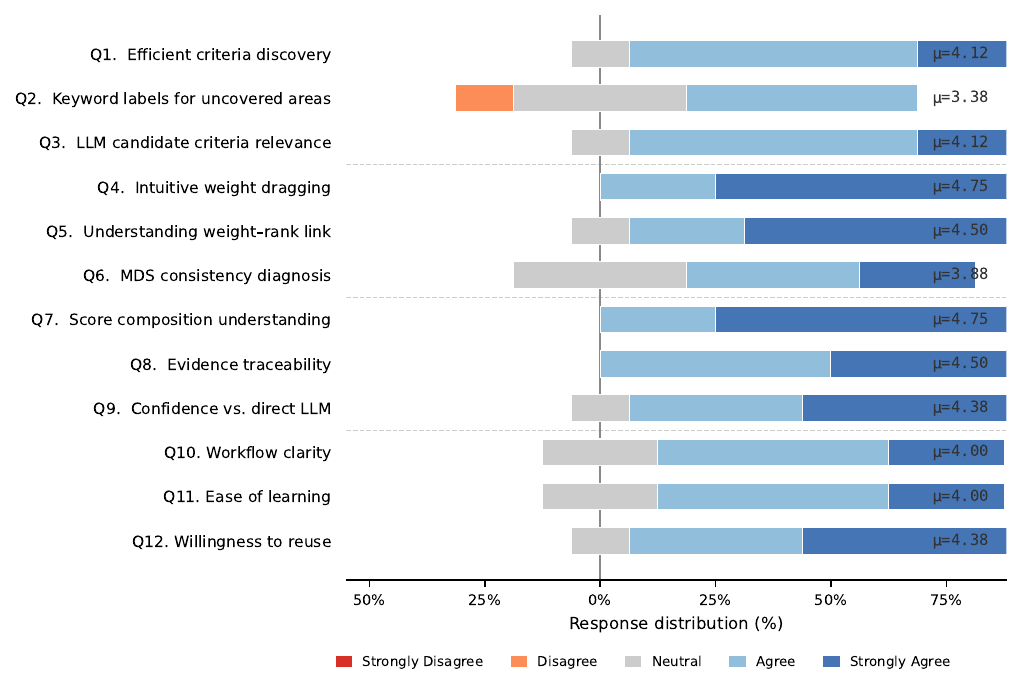}
  \caption{User study Likert scale results ($n=8$, 5-point scale). Bar charts are center-aligned at the neutral option, with negative ratings on the left and positive on the right. Dashed lines separate the four evaluation dimensions.}
  \label{fig:likert}
\end{figure}

Fig.~\ref{fig:likert} shows the results. In the \emph{criteria discovery} dimension, Q1 ($\mu=4.13$) and Q3 ($\mu=4.13$) received positive ratings, with participants finding the system helped them identify evaluation dimensions more efficiently. Q2 ($\mu=3.38$) received the lowest score, reflecting insufficient interpretability of scatter plot keyword labels---P2 mentioned ``keywords are often just one or two words; I can only guess.'' The \emph{weight adjustment} dimension was rated positively overall (Q4 $\mu=4.75$, Q5 $\mu=4.50$). Q6 (MDS consistency diagnosis, $\mu=3.88$) showed polarization: participants with AHP background gave high scores, while those with less background found it abstract. The \emph{evidence traceability} dimension received the highest ratings (Q7 $\mu=4.75$, Q8 $\mu=4.50$, Q9 $\mu=4.38$); P1 commented: ``The system's interactive discovery approach is more convincing than directly getting results from an LLM.'' P8 considered the decision report ``could be used directly as a buying guide.'' For \emph{overall usability}, the workflow was considered clear (Q10 $\mu=4.00$), the system easy to learn (Q11 $\mu=4.00$), and participants expressed willingness to use it in future decisions (Q12 $\mu=4.38$).

\subsection{Quantitative Analysis: Consistency Repair Effectiveness}
\label{sec:eval:quantitative}

To verify \sys{}'s consistency repair mechanism (\S\ref{sec:method:weight}), we designed an automated experiment: using 20 cross-domain decision objectives (spanning electronics, travel, dining, education, health, finance, etc.), we called the LLM to generate complete AHP hierarchies and pairwise comparison matrices for each. Each tree contains a top-level matrix and sub-matrices under each top-level criterion, producing 117 comparison matrices in total. For each matrix, we recorded the LLM's directly generated initial consistency ratio $\mathrm{CR}_{\mathrm{init}}$, then ran the discrete local search repair algorithm (\S\ref{sec:method:weight}) and recorded the repaired $\mathrm{CR}_{\mathrm{rep}}$.

\begin{figure}[t]
  \centering
  \includegraphics[width=0.8\linewidth]{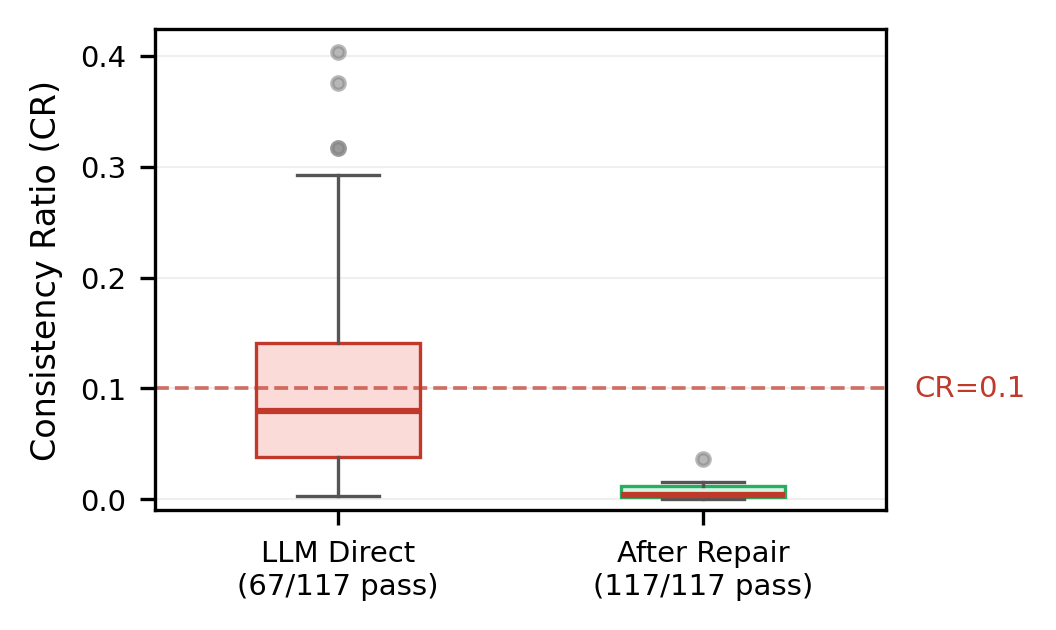}
  \caption{Comparison of CR distributions between LLM-generated and repaired matrices. The red dashed line marks the $\mathrm{CR} = 0.1$ threshold.}
  \label{fig:cr_boxplot}
\end{figure}

Results are shown in Fig.~\ref{fig:cr_boxplot}. Of the 117 LLM-generated matrices, only 57.3\% (67/117) satisfied $\mathrm{CR} \leq 0.1$ (mean 0.105, std 0.084, max 0.404). After repair, the pass rate reached 100\%, with mean CR dropping to 0.007 (std 0.006, max 0.037). This result demonstrates that LLM-generated pairwise comparisons lack transitive consistency and carry risk if used directly; the discrete local search effectively projects inconsistent matrices into the feasible region while preserving similarity to original judgments through log-distance minimization.

\section{Discussion}
\label{sec:discussion}

After system completion, we presented the full functionality to our three collaborating experts (E1--E3, \S\ref{sec:req:background}) and collected in-depth feedback. The following combines expert feedback and user study results to discuss design implications and limitations.

\subsection{Expert Feedback and Design Implications}

\textbf{Value and boundaries of structured frameworks.} \sys{}'s design embodies a judgment: in deliberative decisions, an explicit criteria-weight-evidence structure is more valuable than end-to-end LLM output---even at higher interaction cost. E1 noted that the AHP framework gives weight adjustment a mathematical foundation: ``every user adjustment is traceable and explainable.'' The evidence traceability dimension received the highest user study ratings (Q7--Q9 means $\geq$4.38), confirming the critical role of transparency in decision confidence. However, P6 pointed out that for lightweight purchases, multi-turn LLM dialogue is more efficient. This feedback delineates the system's applicability boundary: when decision outcomes must be justified to others (e.g., corporate procurement reports) or involve multi-party trade-offs (e.g., team negotiation), the structured decision process serves not only as an analytical tool but also as a vehicle for communication and persuasion.

\textbf{Prior knowledge threshold for consistency diagnosis.} E1 particularly praised the Consistency Diagnosis View: ``In traditional AHP, users have no idea where inconsistency comes from. The MDS space lets me `see' contradictions for the first time.'' However, this view showed polarization in the user study (Q6 $\mu=3.88$): participants with AHP background gave high scores, while those with less background found it abstract. Future work should design progressive guidance to lower the learning barrier.

\textbf{Coverage and iteration termination.} Review text naturally contains decision-irrelevant content (e.g., pure narratives, logistics experiences), so coverage has a data-determined ceiling and higher is not always better. In the case studies, E2 observed that marginal coverage gains diminished rapidly in later iterations while rankings stabilized, and used this to judge that effective coverage had reached saturation. Future work can combine marginal coverage increments with ranking stability~\cite{amershi2019guidelines} to design clearer iteration termination indicators.

\subsection{Limitations and Future Work}

The \sys{} framework is not limited to consumer product selection---supplier evaluation, project review, and other scenarios requiring extraction of evaluation dimensions from unstructured text can also benefit, and the core mechanism does not depend on pre-existing numerical ratings (the LLM can generate criterion-level ratings directly). However, the current system has several limitations: (1) criterion-level scores are based on the mean of snippet source review ratings, which may be unreliable when snippet counts are small; (2) the system adopts AHP as its decision framework, whose hierarchical structure assumes independence among criteria; when significant dependencies exist between criteria (e.g., ``noise level'' affecting ``sleep quality''), methods such as ANP (Analytic Network Process) that support feedback and dependencies among criteria are needed, though the corresponding visualization design and consistency diagnosis mechanisms would be considerably more complex; (3) the system lacks spatial exploration; geographically sensitive decisions would benefit from map component integration.

Future plans include: adding convergence assessment metrics to quantify the maximum potential impact of remaining blind spots on rankings; supporting multi-user collaborative decision making; and exploring generalization to non-review text types.

\section{Conclusion}
\label{sec:conclusion}

We present \sys{}, a visual analytics system for multi-criteria decision making from review text. The core design philosophy is to bring dimension discovery itself---not merely dimension manipulation---into the visual analytics loop: the LLM-generated model provides a starting point but not an endpoint, coverage gaps in the embedding space give users intuitive visual anchors to guide exploration, and AHP consistency constraints ensure mathematical soundness throughout iteration. Two case studies verify the system's ability to discover long-tail dimensions missed by the initial model, an eight-participant user study shows positive ratings across criteria discovery, weight adjustment, evidence traceability, and overall usability, and a quantitative experiment confirms that the consistency repair mechanism raises the CR pass rate of LLM-generated matrices from 57.3\% to 100\%.


%% file: template.bib
@article{svoboda2024ahp,
  title={Enhancing multi-criteria decision analysis with AI: Integrating analytic hierarchy process and GPT-4 for automated decision support},
  author={Svoboda, Igor and Lande, Dmytro},
  journal={arXiv preprint arXiv:2402.07404},
  year={2024}
}

@article{vahidnia2025multiagent,
  title={Multi-agent systems of large language models as weight assigners: An approach to collaborative weighting in spatial multi-criteria decision-making},
  author={Vahidnia, Mohammad H},
  journal={Geomatica},
  pages={100071},
  year={2025},
  publisher={Elsevier}
}

@article{park2025ahp,
  title={Enhancing Analytic Hierarchy Process Modelling Under Uncertainty With Fine-Tuning LLM},
  author={Park, Haeun and Oh, Hyunjoo and Gao, Feng and Kwon, Ohbyung},
  journal={Expert Systems},
  volume={42},
  number={6},
  pages={e70051},
  year={2025},
  publisher={Wiley Online Library}
}

@article{wang2025oneforall,
  title={One for all: A general framework of LLMs-based multi-criteria decision making on human expert level},
  author={Wang, Hui and Zhang, Fafa and Mu, Chaoxu},
  journal={arXiv preprint arXiv:2502.15778},
  year={2025}
}

@article{wu2026doc2ahp,
  title={Doc2AHP: Inferring Structured Multi-Criteria Decision Models via Semantic Trees with LLMs},
  author={Wu, Hongjia and Zhou, Shuai and Zhang, Hongxin and Chen, Wei},
  journal={arXiv preprint arXiv:2601.16479},
  year={2026}
}

@inproceedings{lu2024ahpllm,
  title={AHP-powered LLM reasoning for multi-criteria evaluation of open-ended responses},
  author={Lu, Xiaotian and Li, Jiyi and Takeuchi, Koh and Kashima, Hisashi},
  booktitle={Findings of the Association for Computational Linguistics: EMNLP 2024},
  pages={1847--1856},
  year={2024}
}

@inproceedings{mareschal2009gaia,
  title={Visual PROMETHEE: Developments of the PROMETHEE \& GAIA multicriteria decision aid methods},
  author={Mareschal, Bertrand and De Smet, Yves},
  booktitle={2009 IEEE International conference on industrial engineering and engineering management},
  pages={1646--1649},
  year={2009},
  organization={IEEE}
}

@article{ishizaka2016gaia,
  title={Which energy mix for the UK (United Kingdom)? An evolutive descriptive mapping with the integrated GAIA (graphical analysis for interactive aid)--AHP (analytic hierarchy process) visualization tool},
  author={Ishizaka, Alessio and Siraj, Sajid and Nemery, Philippe},
  journal={Energy},
  volume={95},
  pages={602--611},
  year={2016},
  publisher={Elsevier}
}

@article{asahi1995treemap,
  title={Using treemaps to visualize the analytic hierarchy process},
  author={Asahi, Toshiyuki and Turo, David and Shneiderman, Ben},
  journal={Information Systems Research},
  volume={6},
  number={4},
  pages={357--375},
  year={1995},
  publisher={INFORMS}
}

@article{susmaga2024topsis,
  title={Towards explainable TOPSIS: Visual insights into the effects of weights and aggregations on rankings},
  author={Susmaga, Robert and Szcz{\k{e}}ch, Izabela and Brzezinski, Dariusz},
  journal={Applied Soft Computing},
  volume={153},
  pages={111279},
  year={2024},
  publisher={Elsevier}
}

@article{sun2023fuzzyahp,
  title={A visual analytics approach for multi-attribute decision making based on intuitionistic fuzzy AHP and UMAP},
  author={Sun, Yan and Zhou, Xiaojun and Yang, Chunhua and Huang, Tingwen},
  journal={Information Fusion},
  volume={96},
  pages={269--280},
  year={2023},
  publisher={Elsevier}
}

@article{gratzl2013lineup,
  title={Lineup: Visual analysis of multi-attribute rankings},
  author={Gratzl, Samuel and Lex, Alexander and Gehlenborg, Nils and Pfister, Hanspeter and Streit, Marc},
  journal={IEEE transactions on visualization and computer graphics},
  volume={19},
  number={12},
  pages={2277--2286},
  year={2013},
  publisher={IEEE}
}

@inproceedings{carenini2004valuecharts,
  title={Valuecharts: analyzing linear models expressing preferences and evaluations},
  author={Carenini, Giuseppe and Loyd, John},
  booktitle={Proceedings of the working conference on Advanced visual interfaces},
  pages={150--157},
  year={2004}
}

@article{pajer2017weightlifter,
  title={Weightlifter: Visual weight space exploration for multi-criteria decision making},
  author={Pajer, Stephan and Streit, Marc and Torsney-Weir, Thomas and Spechtenhauser, Florian and M{\"o}ller, Torsten and Piringer, Harald},
  journal={IEEE transactions on visualization and computer graphics},
  volume={23},
  number={1},
  pages={611--620},
  year={2016},
  publisher={IEEE}
}

@inproceedings{schmid2022rankasco,
  title={RankASco: A Visual Analytics Approach to Leverage Attribute-Based User Preferences for Item Rankings},
  author={Schmid, Jenny and Cibulski, Lena and Al Hazwani, Ibrahim and Bernard, J{\"u}rgen},
  booktitle={EuroVA@ EuroVis},
  pages={7--11},
  year={2022}
}

@article{wall2018podium,
  title={Podium: Ranking data using mixed-initiative visual analytics},
  author={Wall, Emily and Das, Subhajit and Chawla, Ravish and Kalidindi, Bharath and Brown, Eli T and Endert, Alex},
  journal={IEEE transactions on visualization and computer graphics},
  volume={24},
  number={1},
  pages={288--297},
  year={2017},
  publisher={IEEE}
}

@inproceedings{chang2020gaggle,
  title={Gaggle: Visual Analytics using Interactive Multi-Model Steering},
  author={Das, Subhajit and Cashman, Dylan and Chang, Remco and Endert, Alex},
  booktitle = {Graphics Interface},
  year = {2020}
}

@inproceedings{odonovan2008peerchooser,
  title={PeerChooser: visual interactive recommendation},
  author={O'Donovan, John and Smyth, Barry and Gretarsson, Brynjar and Bostandjiev, Svetlin and H{\"o}llerer, Tobias},
  booktitle={Proceedings of the SIGCHI Conference on Human Factors in Computing Systems},
  pages={1085--1088},
  year={2008}
}

@article{choo2018visirr,
  title={VisIRR: A visual analytics system for information retrieval and recommendation for large-scale document data},
  author={Choo, Jaegul and Kim, Hannah and Clarkson, Edward and Liu, Zhicheng and Lee, Changhyun and Li, Fuxin and Lee, Hanseung and Kannan, Ramakrishnan and Stolper, Charles D and Stasko, John and others},
  journal={ACM Transactions on Knowledge Discovery from Data (TKDD)},
  volume={12},
  number={1},
  pages={1--20},
  year={2018},
  publisher={ACM New York, NY, USA}
}

@article{becker2024graphical,
  title={Graphical analysis of consistency in the AHP/ANP pairwise comparison matrix of criteria or decision alternatives},
  author={Becker, Jaros{\l}aw and Becker, Aneta},
  journal={Procedia Computer Science},
  volume={246},
  pages={4805--4814},
  year={2024},
  publisher={Elsevier}
}

@article{chen2024bge,
  title={Density-based clustering based on hierarchical density estimates},
  author={Campello, Ricardo JGB and Moulavi, Davoud and Sander, J{\"o}rg},
  booktitle={Pacific-Asia conference on knowledge discovery and data mining},
  pages={160--172},
  year={2013},
  organization={Springer}
}

@article{mcinnes2018umap,
  title={Umap: Uniform manifold approximation and projection for dimension reduction},
  author={McInnes, Leland and Healy, John and Melville, James},
  journal={arXiv preprint arXiv:1802.03426},
  year={2018}
}

@inproceedings{campello2013hdbscan,
  title={Density-based clustering based on hierarchical density estimates},
  author={Campello, Ricardo JGB and Moulavi, Davoud and Sander, J{\"o}rg},
  booktitle={Pacific-Asia conference on knowledge discovery and data mining},
  pages={160--172},
  year={2013},
  organization={Springer}
}

@inproceedings{hutto2014vader,
  title={Vader: A parsimonious rule-based model for sentiment analysis of social media text},
  author={Hutto, Clayton and Gilbert, Eric},
  booktitle={Proceedings of the international AAAI conference on web and social media},
  volume={8},
  number={1},
  pages={216--225},
  year={2014}
}

@inproceedings{pirolli2005sensemaking,
  title={The sensemaking process and leverage points for analyst technology as identified through cognitive task analysis},
  author={Pirolli, Peter and Card, Stuart},
  booktitle={Proceedings of international conference on intelligence analysis},
  volume={5},
  number={1},
  pages={2--4},
  year={2005},
  organization={McLean, VA, USA}
}

@article{endert2012semantic,
  title={Semantic interaction for sensemaking: inferring analytical reasoning for model steering},
  author={Endert, Alex and Fiaux, Patrick and North, Chris},
  journal={IEEE Transactions on Visualization and Computer Graphics},
  volume={18},
  number={12},
  pages={2879--2888},
  year={2012},
  publisher={IEEE}
}

@article{sacha2017human,
  title={What you see is what you can change: Human-centered machine learning by interactive visualization},
  author={Sacha, Dominik and Sedlmair, Michael and Zhang, Leishi and Lee, John A and Peltonen, Jaakko and Weiskopf, Daniel and North, Stephen C and Keim, Daniel A},
  journal={Neurocomputing},
  volume={268},
  pages={164--175},
  year={2017},
  publisher={Elsevier}
}

@article{wu2010opinionseer,
  title={OpinionSeer: interactive visualization of hotel customer feedback},
  author={Wu, Yingcai and Wei, Furu and Liu, Shixia and Au, Norman and Cui, Weiwei and Zhou, Hong and Qu, Huamin},
  journal={IEEE transactions on visualization and computer graphics},
  volume={16},
  number={6},
  pages={1109--1118},
  year={2010},
  publisher={IEEE}
}

@article{liu2019bridging,
  title={Bridging text visualization and mining: A task-driven survey},
  author={Liu, Shixia and Wang, Xiting and Collins, Christopher and Dou, Wenwen and Ouyang, Fangxin and El-Assady, Mennatallah and Jiang, Liu and Keim, Daniel A},
  journal={IEEE transactions on visualization and computer graphics},
  volume={25},
  number={7},
  pages={2482--2504},
  year={2018},
  publisher={IEEE}
}

@article{park2018conceptvector,
  title={ConceptVector: Text visual analytics via interactive lexicon building using word embedding},
  author={Park, Deokgun and Kim, Seungyeon and Lee, Jurim and Choo, Jaegul and Diakopoulos, Nicholas and Elmqvist, Niklas},
  journal={IEEE transactions on visualization and computer graphics},
  volume={24},
  number={1},
  pages={361--370},
  year={2017},
  publisher={IEEE}
}

@article{zhao2025leva,
  title={Leva: Using large language models to enhance visual analytics},
  author={Zhao, Yuheng and Zhang, Yixing and Zhang, Yu and Zhao, Xinyi and Wang, Junjie and Shao, Zekai and Turkay, Cagatay and Chen, Siming},
  journal={IEEE transactions on visualization and computer graphics},
  volume={31},
  number={3},
  pages={1830--1847},
  year={2024},
  publisher={IEEE}
}

@article{shen2023nlvis,
  title={Towards natural language interfaces for data visualization: A survey},
  author={Shen, Leixian and Shen, Enya and Luo, Yuyu and Yang, Xiaocong and Hu, Xuming and Zhang, Xiongshuai and Tai, Zhiwei and Wang, Jianmin},
  journal={IEEE transactions on visualization and computer graphics},
  volume={29},
  number={6},
  pages={3121--3144},
  year={2022},
  publisher={IEEE}
}

@article{zhang2023absa,
  title={A survey on aspect-based sentiment analysis: Tasks, methods, and challenges},
  author={Zhang, Wenxuan and Li, Xin and Deng, Yang and Bing, Lidong and Lam, Wai},
  journal={IEEE Transactions on Knowledge and Data Engineering},
  volume={35},
  number={11},
  pages={11019--11038},
  year={2022},
  publisher={IEEE}
}

@article{crawford1985llsm,
  title={A note on the analysis of subjective judgment matrices},
  author={Crawford, Gordon and Williams, Cindy},
  journal={Journal of mathematical psychology},
  volume={29},
  number={4},
  pages={387--405},
  year={1985},
  publisher={Elsevier}
}

@article{sparckjones1972idf,
  title={A statistical interpretation of term specificity and its application in retrieval},
  author={Sparck Jones, Karen},
  journal={Journal of documentation},
  volume={28},
  number={1},
  pages={11--21},
  year={1972},
  publisher={MCB UP Ltd}
}

@book{borg2005mds,
  title={Modern multidimensional scaling: Theory and applications},
  author={Borg, Ingwer and Groenen, Patrick JF},
  year={2005},
  publisher={Springer}
}

@article{bostock2011d3,
  title={D$^3$ data-driven documents},
  author={Bostock, Michael and Ogievetsky, Vadim and Heer, Jeffrey},
  journal={IEEE transactions on visualization and computer graphics},
  volume={17},
  number={12},
  pages={2301--2309},
  year={2011},
  publisher={IEEE}
}

@book{ericsson1993protocol,
  author = {K. A. Ericsson and H. A. Simon},
  title = {Protocol Analysis: Verbal Reports as Data},
  publisher = {MIT Press},
  edition = {revised},
  year = {1993}
}

@article{dimara2022decision,
  title={A critical reflection on visualization research: Where do decision making tasks hide?},
  author={Dimara, Evanthia and Stasko, John},
  journal={IEEE Transactions on Visualization and Computer Graphics},
  volume={28},
  number={1},
  pages={1128--1138},
  year={2021},
  publisher={IEEE}
}

@article{gleicher2011comparison,
  title={Visual comparison for information visualization},
  author={Gleicher, Michael and Albers, Danielle and Walker, Rick and Jusufi, Ilir and Hansen, Charles D and Roberts, Jonathan C},
  journal={Information Visualization},
  volume={10},
  number={4},
  pages={289--309},
  year={2011},
  publisher={SAGE Publications Sage UK: London, England}
}

@article{endert2017integrating,
  title={The state of the art in integrating machine learning into visual analytics},
  author={Endert, Alex and Ribarsky, William and Turkay, Cagatay and Wong, BL William and Nabney, Ian and Blanco, I D{\'\i}az and Rossi, Fabrice},
  booktitle={Computer Graphics Forum},
  volume={36},
  number={8},
  pages={458--486},
  year={2017},
  organization={Wiley Online Library}
}

@article{elassady2018progressive,
  title={Progressive learning of topic modeling parameters: A visual analytics framework},
  author={El-Assady, Mennatallah and Sevastjanova, Rita and Sperrle, Fabian and Keim, Daniel and Collins, Christopher},
  journal={IEEE transactions on visualization and computer graphics},
  volume={24},
  number={1},
  pages={382--391},
  year={2017},
  publisher={IEEE}
}

@inproceedings{boggust2022embedding,
  title={Embedding comparator: Visualizing differences in global structure and local neighborhoods via small multiples},
  author={Boggust, Angie and Carter, Brandon and Satyanarayan, Arvind},
  booktitle={Proceedings of the 27th international conference on intelligent user interfaces},
  pages={746--766},
  year={2022}
}

@inproceedings{amershi2019guidelines,
  title={Guidelines for human-AI interaction},
  author={Amershi, Saleema and Weld, Dan and Vorvoreanu, Mihaela and Fourney, Adam and Nushi, Besmira and Collisson, Penny and Suh, Jina and Iqbal, Shamsi and Bennett, Paul N and Inkpen, Kori and others},
  booktitle={Proceedings of the 2019 chi conference on human factors in computing systems},
  pages={1--13},
  year={2019}
}

@article{lam2011empirical,
  title={Empirical studies in information visualization: Seven scenarios},
  author={Lam, Heidi and Bertini, Enrico and Isenberg, Petra and Plaisant, Catherine and Carpendale, Sheelagh},
  journal={IEEE transactions on visualization and computer graphics},
  volume={18},
  number={9},
  pages={1520--1536},
  year={2011},
  publisher={IEEE}
}

@article{zheng2023llmjudge,
  title={Judging llm-as-a-judge with mt-bench and chatbot arena},
  author={Zheng, Lianmin and Chiang, Wei-Lin and Sheng, Ying and Zhuang, Siyuan and Wu, Zhanghao and Zhuang, Yonghao and Lin, Zi and Li, Zhuohan and Li, Dacheng and Xing, Eric and others},
  journal={Advances in neural information processing systems},
  volume={36},
  pages={46595--46623},
  year={2023}
}

@article{sacha2019vis4ml,
  title={Vis4ml: An ontology for visual analytics assisted machine learning},
  author={Sacha, Dominik and Kraus, Matthias and Keim, Daniel A and Chen, Min},
  journal={IEEE transactions on visualization and computer graphics},
  volume={25},
  number={1},
  pages={385--395},
  year={2018},
  publisher={IEEE}
}

@inproceedings{wei2022cot,
  title={Chain-of-thought prompting elicits reasoning in large language models},
  author={Wei, Jason and Wang, Xuezhi and Schuurmans, Dale and Bosma, Maarten and Xia, Fei and Chi, Ed and Le, Quoc V and Zhou, Denny and others},
  journal={Advances in neural information processing systems},
  volume={35},
  pages={24824--24837},
  year={2022}
}

@inproceedings{ni2019justifying,
  title={Justifying recommendations using distantly-labeled reviews and fine-grained aspects},
  author={Ni, Jianmo and Li, Jiacheng and McAuley, Julian},
  booktitle={Proceedings of the 2019 conference on empirical methods in natural language processing and the 9th international joint conference on natural language processing (EMNLP-IJCNLP)},
  pages={188--197},
  year={2019}
}

@inproceedings{antognini2020hotelrec,
  title={Hotelrec: a novel very large-scale hotel recommendation dataset},
  author={Antognini, Diego and Faltings, Boi},
  booktitle={Proceedings of the Twelfth Language Resources and Evaluation Conference},
  pages={4917--4923},
  year={2020}
}
